# Magnetic Resonance Imaging Virtual Liver Biopsy Using Radiomics Analysis for the Assessment of Chronic Liver Disease

## Author Names and Degrees:

Jiqing Huang, PhD,[1] Benjamin Leporq, PhD,[1] , Valérie Hervieu, MD,[2] , *Sophie Gaillard* , MS[1], Jerome Dumortier, MD, [3],Olivier Beuf, PhD,[1*],Hélène Ratiney, PhD,[1]

## Author Affiliations:

[1] CREATIS UMR 5220, U1294, Univ Lyon, INSA-Lyon, Université Claude Bernard Lyon 1, UJM-Saint Etienne, CNRS, Inserm; F-69XXX, LYON, France

[2] Department of Ana-tomo-pathology, CHU Edouard Herriot, Hospices Civils de Lyon; F-69003, Lyon, France

[3] Department of Hepatology, CHU Edouard Herriot, Hospices Civils de Lyon; F-69003, Lyon, France

## Corresponding Author Info:

Olivier Beuf

**Address: O.B. INSA LYON Bâtiment Léonard de Vinci, 21 Avenue Jean Capelle, 69621 Villeurbanne Cedex, FRANCE*

*telephone: +33472438520*

*e-mail address: olivier.beuf@creatis.insa-lyon.fr*

## Acknowledgement:

We warmly thank professor Pierre-Jean Valette (radiologist at ''Hospices Civils de Lyon") for his involvement and for facilitating the implementation of this work.

## Grant Support:

This work was supported by the China Scholarship Council (CSC), and LabEx PRIMES (ANR-11-LABX-0063) of Université de Lyon, within the program "Investissements d'Avenir" (ANR-11-IDEX-0007) operated by the French National Research Agency (ANR).

*Abstract—*

**Objectives:** The role of advanced diffusion-weighted imaging (DWI) in chronic liver disease (CLD) has not been fully studied. Chronic liver disease (CLD) is a progressive deterioration of liver functions, caused by one or more etiology. This study was aimed to investigate whether radiomics features extracted from individual or combined magnetic resonance imaging sequences, such as T1-weighted, T2-weighted images, or quantitative maps from chemical shift encoded, diffusion-weighted imaging, can effectively classify inflammation and fibrosis in CLD.

**Method:** Seventy-seven patients with CLD were enrolled in this study. Each participant underwent both MRI examinations and liver biopsy. The biopsy procedure was applied to quantitatively or semi-qualitatively analyze several histology features, steatosis, inflammation, and fibrosis. Radiomic features were extracted, selected, reduced, and used to train the inflammation and fibrosis classification based on random forest models. The performances of classifiers were evaluated by the receiver operating characteristic curve (ROC), accuracy, precision, sensitivity, specificity, and DeLong tests.

**Result: The random forest model** achieved the area under the curve (AUC) of 0.85 and 0.86 for inflammation and fibrosis classification, respectively.

**Conclusion:** This study demonstrated that the MRI-based radiomics features hold potential in the inflammation and fibrosis classification.

## INTRODUCTION

Chronic liver disease (CLD) is a leading cause of morbidity and mortality worldwide, accounting for an estimated two million deaths per year [1]. It includes a large variety of hepatic diseases that, in most subtypes, damage the liver in similar ways [2]. CLD progression may be characterized by histology with the presence of steatosis, inflammation, fibrosis, and iron overload regardless of the underlying etiologies [3]. While liver biopsy is a prerequisite to histology, there is still an unmet medical requirement to characterize CLD in a non-invasive way [3]. Among them, inflammation and fibrosis have emerged as two pivotal factors in that inflammation plays a role in the early liver fibrosis process [4], while fibrosis affects the CLD prognosis and treatment strategy [5].

To date, ultrasound (US) and Magnetic resonance imaging (MRI) are used for CLD diagnosis [6]. Unfortunately, conventional imaging techniques are limited in assisting in the diagnosis of advanced CLD [6]. Transient elastography (Fibroscan) based on US and magnetic resonance elastography (MRE), quantitative elastic properties of tissues related to fibrosis and inflammation can be estimated [7]. However, reliable Fibroscan or MRE examination requires an external device (to generate mechanical waves), experienced physicians and operators, and careful analysis of fat content and iron overload [8]. In practice, diagnostic performance is intricately linked to the physician's expertise and experience [8]. Moreover, elastography is considered irrelevant in many studies for patients with early fibrosis and unreliable in overweight patients [9].

In the literature, extensive evidence demonstrated the value of multiple MRI techniques in the assessment of fibrosis or inflammation [10–16]. The T1-weighted (T1WI) and T2-weighted (T2WI) sequences may reflect disordered structures or microscopic changes in the tissue [10, 11]. T2*-mapping is able to evaluate the magnetic field inhomogeneities caused by iron overload [12]; proton density fat fraction (PDFF) provides information about steatosis and fat content [13]; Diffusion-weighted imaging (DWI) has the capability to assess fibrosis and is potentially associated with inflammation [14–16]. Collectively, MRI has been considered a promising imaging modality for the diagnosis of CLD, thanks to multi-contrast and high spatial resolution and quantitative protocols.

In addition, an emerging discipline known as radiomics has garnered growing attention in the characterization of CLD [17, 18]. The meaningful features are extracted, selected, and then used to assist in characterizing various histological features. In this study, we explored the performance of radiomics analysis for the evaluation of fibrosis and inflammation in CLD using images derived from conventional and advanced MRI acquisitions.

## MATERIALS AND METHODS

### *A. Patients*

This study protocol was approved by the local ethics committee (gCCP 2013-A00568-39). Seventy-seven patients (33 women, 44 men, mean age 48.9 ± 15.0 years; mean weight: 77.1 ± 16.6 Kg; mean height: 1.67 ± 0.09 m) with histology-proven CLD diseases were enrolled. The patients have one or more etiologies including viral hepatitis B, and C (VHB, n = 20, and VHC, n = 17); Non-alcoholic steatohepatitis (NASH, n = 19); and others (cholangiopathy; vascular diseases; steatosis; immune disorder; hemochromatosis, n = 17). All patients underwent biopsy and MRI acquisition with information and consent.

### *B. Liver biopsy*

Biopsies were done using an 18 or 16 G biopsy needle via a transparietal approach to collect tissue samples. After being fixed by 10% neutral buffered formalin, the tissues were embedded in paraffin. The small portion of the liver was sectioned 4mm thick and was mounted on glass slides. Each section was stained with hematoxylin-eosin-saffron, iron stain, and Masson trichrome reagents. The histopathological evaluation was co-evaluated by two pathologists (VH) with more than 10 years of experience. The pathologists were blinded to any clinical information and MRI results. The steatosis and fibrosis percentages were calculated by the ratio of components on the slide with computerized morphometry. The inflammation was analyzed separately in the lesion region (portal or lobular) considering their respective degrees (portal: P=0: no, P=1: mild, P=2: severe; lobular: L=0: no, L=1: mild, L=2: severe).

The steatosis was classified into four grades with Brunt's score (S0: 0-5%, S1: 5-33%, S2: 33-66%, S3 >66%). The inflammation was classified into 3 grades (A0-A2, grade= max (P, L)). The fibrosis was stratified using the ISHAK score. The score is mainly related to fibrosis, but it is also related to the activity and presence of other associated lesions (F0: no fibrosis, F1: some portal areas ± short fibrosis septa, F2: most portal areas ± short fibrosis septa, F3: most portal areas + occasional portal to portal bridging, F4： most portal areas + marked bridging, F5: marked bridging + occasional nodules, F6: cirrhosis, probable or define).

The binary classification labels were set for fibrosis and inflammation (non-significant fibrosis [F0–F2] vs. significant fibrosis [F3–F6], and no inflammation [A0] vs. hepatitis [A1–A2], respectively).

### *C. Image acquisition*

The acquisition protocol included conventional T1WI with contrast agent gadobentate-dimeglumine (Gd-BOPTA) injection, T2WI, chemical shift encoded (CSE-MRI) - providing proton density fat fraction (PDFF) and T2*maps - and multiple b-value diffusion-weighted imaging (DWI). All MRI examinations were carried out at 3-Tesla (GE MR750, GE Healthcare, Chicago, IL, USA) with slices oriented in the transverse plane. The MRI protocols consisted of 3D fat-suppressed LAVA FLEX T1WI sequence (TE/TRE =3.4/4.2 ms, acquired slice thickness/ reconstructed slice thickness = 3.4/1.7 mm, field of view = 380× 380 $mm^2$, flip angle = 15º, acquisition matrix = 320 × 224, reconstruction matrix = 512×512) and turbo spin echo T2WI PROPELLER sequence (TR/TE = 3157.9/113.1 ms, slice thickness = 4.0 mm, field of view = 400× 400 $mm^2$, flip angle = 110 º, acquisition matrix = 384× 384, reconstruction matrix = 512×512).

Chemical shift-encoded multiple spoiled gradients recalled echo (SPGR) sequence with 4 echoes (TR/ TE = 100/ TE0: 1.1, TE1: 2.3, TE2: 3.6, TE3: 4.8 ms, acquired slice thickness/ reconstructed slice thickness = 10.0/5.0 mm, field of view = 410× 410 mm2, flip angles = 5º, 15º, 30º, 45º, acquisition matrix = 128 × 128, reconstruction matrix = 256×256) was used to compute PDFF and T2*-maps [19].

DWI was performed with spin-echo echo planar imaging (SE-EPI) enhanced diffusion-weighted imaging (TR/ TE = 2050.0/ 54.3 ms, slice thickness/spacing between slices =8.0/4.0 mm, field of view = 400 × 300 $mm^2$, flip angle = 90º, acquisition matrix = 96 × 128, reconstruction matrix = 256×256). Twelve b-values (0, 10, 20, 40, 60, 80, 100, 200, 300, 400, 600, 800 s/ $mm^2$) were acquired with a tetrahedral diffusion gradient encoding. The diffusion models including the mono-exponential model (MONO), intravoxel incoherent motion (IVIM), stretched exponential model (SEM), statistical diffusion model (Stat_D), were fitted with nonlinear least square (NLS, for MONO), and Bayesian shrinkage approach with NLS initialization (for others model) in MATLAB 2019b (The MathWorks Inc., Natick, MA, USA) [16]. The model and corresponding parameters are listed in TABLE I.

The whole liver was manually segmented in two dimensions from a single slice located in the middle of the liver by J.H. and validated by B.L.

### *D. Radiomic feature extraction*

Radiomic feature extraction was carried out using the Pyradiomics toolbox (in Python version 3.7) [20]. For preprocessing, all MRI parameters including T1WI, T2WI, b=0 diffusion image, PDFF, T2*-maps, and diffusion parametric maps (eight maps, listed in TABLE I) were normalized with z-score. Then, ninety-four features (first-order statistics: 19, gray level co-occurrence matrix: 24, gray level run length matrix: 16, gray level size zone matrix neighboring: 16, gray-tone difference matrix: 5, gray level dependence matrix: 14) were extracted from original normalized images and wavelet decomposed images (eight decompositions). As a result, a total of 94 × 13 × 9 = 10 998 features were extracted for each patient.

### *E. Feature selection and reduction*

To avoid the curse of dimensionality problem, a features reduction method has been implemented using scikit-learn. First, unsupervised principal component analysis (PCA) was adopted to select sixty features with low redundancy features per contrast. Second, for each classification task, (inflammation or significant fibrosis), the supervised Fisher method was used to discard up to 20 features with both small intra-class differences and large inter-class differences per modality. Finally, the desired features contributing to classification were determined by the support vector machines-recursive feature elimination (SVM-REF) method. The maximal number of features was set to 15 for inflammation classification, and to 20 for fibrosis classification in single and multiple sequences model construction. For the inflammation classification task, the dataset was resampled by adaptive synthetic (ADASYN) algorithm to balance the numbers of the patients in two groups (severe vs. mild patients: before ADASYN 54:23; after ADASYN 56:54). To properly adapt to the classification task, the feature numbers and modalities combination were iteratively tuned from a few to more.

### *F. Model constructions*

Patients were divided into training (75%) and test sets (25%) according to inflammation grades or fibrosis grades by stratified strategy. Random Forest Classifier (RFC) with scikit-learn was constructed on the training data with stratified k-folders cross-validation (k=5). The optimized classifier with minimal features and modalities combination was determined on the training dataset. For CSE and diffusion sequences, the RFC was trained with one or more features extracted from the parametric maps. In each combination, grid searching was used to find optimal hyperparameters. The best RFC with optimized hyperparameters was retraining in k-folders and k classifiers were built. Then the performance of k classifiers was both validated on the training and test datasets.

### *G. Statistical analysis and diagnostic performance*

Fisher's exact test and student-t test were used to compare the difference in categorical and continuous variables in the clinical characteristics of patients, respectively. The ROC curves and AUC analysis were plotted to assess and compare the quantitative performance both on the training dataset and test dataset. The cutoff values provided a balanced combination of sensitivity and specificity were calculated on the training dataset. Then, to quantitatively analyze the diagnostic performances of different combinations, the accuracy, precision, sensitivity, and specificity were computed on training and test set at cutoff values. The AUCs of model trained from radiomics extracted from different image combinations were compared by the Delong test.

# RESULTS

*A. Clinical characteristics*

The baseline characteristics of patients are summarized in Table II. As inflammation and fibrosis are distinct processes, they can be distinguished ($p<0.01$) from each other considerably. In our cohort, viral hepatitis constitutes a higher proportion of patients with inflammation.

*B. Comparison of features extracted from a single sequence for inflammation or fibrosis*

For inflammation classification, radiomics extracted from PDFF and a combination of diffusion parameters {f, D*, σ, ADC, B0} provided the best performances: AUC (training set: 0.99; test set: 0.94) compared to other combinations with other sequences. Meanwhile, PDFF and a combination of diffusion parameters {DDC, Diffusion, Ds} achieved an AUC of (training set: 0.98; test set: 0.81) for fibrosis.

For the construction of further multiple sequences, the terms "CSE combination" and "diffusion combination" refer to the parameter combinations in the last sentence. These combinations result in the highest AUC values obtained during single sequence constructions for inflammation or fibrosis. Figure 1 a-d shows the ROC curves of the aforementioned sequence for fibrosis or inflammation in the training and test set. In addition, the cut-off value was determined in the training set. For each classification task, the quantitative metrics (accuracy, precision, sensitivity, specificity) at corresponding cut-off values were calculated as shown in Table III. In the case of the fibrosis multiple combination model, the T1WI sequence was excluded from the construction of the multiple sequences model because its AUC dropped below 0.5.

*C. Diagnostic performances of models training from different radiomics features combination for the classification of liver fibrosis and inflammation*

The RFC was then trained by combining features extracted from different images from two to multiple sequences. ROC curves illustrating the diagnostic performances for the classification of fibrosis and inflammation are shown in Figure 2 a-f and Figure 3 a-d. The performances of different combinations evaluation are provided in Table IV. Besides, a combination of {T1WI, T2WI, CSE, Diffusion} and

{T2WI, CSE, Diffusion} performed best on inflammation (AUC: training 1.00; test 1.00) and fibrosis (AUC: training 0.99; test 0.89) respectively.

Given that inflammation is significantly associated with fibrosis, we also trained an integrated model taking into account the probability score of inflammation predicted by the best inflammation classifier and radiomics features extracted from three MRI sequences (T2WI, CSE, and DWI). The performance of the integrated model was obviously increased than the MRI-only model (AUC training set: 0.99; AUC test set 0.97) with only 14 features (13 MRI features+ inflammation likelihood).

### *D. Comparison of multiple sequences and single sequence models*

The Figure 4 summarizes the Delong test p-values computed to compare the classifiers built with radiomics extracted from a single sequence and from a combination of multiple sequences. In Figure 4 a-b, RFC built with radiomics extracted from diffusion parameters displayed improved performances than others for both inflammation and fibrosis (except the CSE in fibrosis classification). In Figure 4 c-d, the models built with all sequence's combinations provided better diagnostic performances compared to other sequences for both inflammation and fibrosis.

## Discussion

In this study, we investigated radiomics analysis based on features extracted from one or more conventional MRI acquisitions (injected T1 weighted and fat-saturated T2 weighted) and parametric maps such as T2*, PDFF and diffusions maps computed from multiple b-values acquisition with different function models to classify inflammation and fibrosis in patients with CLD. In single sequence model construction, we found that the model built with features extracted from diffusion maps resulted in significantly improved diagnostic performances in comparison with others for both fibrosis and inflammation assessment. When we add information from the combination of multiple sequences, the diagnostic performances were significantly improved. RFC built with multiple sequences or joined with inflammation regression score provided the best classification result for inflammation and fibrosis. The sensitivity and specificity of the multi-sequence model offered outstanding predictive performances and generalization in both the training and testing sets.

A previous work has demonstrated that the diffusion parameters fitted by Bayesian were associated with fibrosis grades. [16]. These previous results are consistent with ours in this present study where diagnostic performances of the models increased when radiomics features extracted from diffusion parameters map were included.

Recently, MRI imaging of inflammation in the liver has been studied. To detect the inflammation at the microscopic level, targeting of immune cells was mostly done with magnetic nanoparticles with T2 MRI applications [21]. More recently, T1WI and T2WI with a high magnetic field (7 Tesla) presented a significant correlation with inflammation in animal study [22]. So far, there is still a lack of validated diagnostic imaging criteria for inflammation in human studies. In this study, the classifiers trained from the combination of T1WI, T2WI, PDFF, T2*, and Diffusion maps, have demonstrated promising performances in classifying key histological features such as fibrosis and inflammation. By comparing the different performances of the models, an RFC built with diffusion parameters provided 93% AUC although it had been shown in previous study that diffusion parameters were not correlated with inflammation [16]. This result may indicate that the radiomics features could capture the histological heterogeneity at microstructure texture. (usually less than 1 cm, [23] ). To our knowledge, it is the first study that successfully classified the liver inflammation grade in patients with multiple etiologies.

In general, the AUC of almost all classifiers built from MRI-only sequences constructed for inflammation was higher than for liver fibrosis classification. The reason could be that fibrosis is a pathology process synergized by multifactorial effects. Mostly, inflammation represented a prominent predisposing factor that triggered an aberrant healing process [24]. However, fibrosis triggered by inflammation is a classical fibrosis development assumption. In literature, fibrosis has been demonstrated that it is also concomitant with iron deposition, and steatosis [25], [26]. As a result, the feature sizes were adaptively set to 15 for inflammation and 20 for fibrosis to adapt to the problem’s complexity. More specifically, the lipotoxicity caused by inflammation and steatosis activates hepatic stellate cells that promote the fibrogenic phenotype and fibrosis [27], [28]. When the liver attempts to regenerate the damaged cells, stellate cells are activated to differentiate into myofibroblasts [29]. Myofibroblasts secreted some proteins to the extracellular environment. With the accumulation of extracellular matrix, scarring tissue begins and replaces the normal tissue in some populations. The process of fibrosis may continue for decades, 10% to 30% of people may have some degree of iron overload [30]. On top of that, it is known that iron overload is an important factor that contributes to fibrosis and could be interesting to take into account in fibrosis classification. Therefore, these complex mechanisms inspired us to build a non-end-to-end model. This cascade classification built with {T2WI, PDFF, T2*, Diffusion sequence, inflammation} obviously improved classifier performance. This combination can be interpreted as follows. T2* could probe iron overload [31]. As the presence of steatosis is a confounding factor for fibrosis evaluation using diffusion MRI, adding its measurement thanks to CSE-derived PDFF (a nowadays recognized imaging marker of steatosis [32]) to diffusion-related parameters could help fibrosis classification. Moreover, evaluated T2WI with liver fibrosis may be associated with an elevated inflammatory component which has been seen in chronic hepatitis and steatohepatitis. The inflammation likelihood offered complementary information. The combination of T2WI feature and inflammation index making it beneficial for fibrosis identification [33]. Overall, the final joint model has used all known fibrosis-related features.

There were however some limitations to our study. First, the dataset was not fully balanced for both inflammation and fibrosis. Especially for inflammation, we had to resample the data for inflammation, which could result in overly optimistic classification results, but gathering this kind of data, including a biopsy, is no easy task. Secondly, if the pan-etiologies aspect of our cohort is interesting since it may increase the model robustness regarding chronic liver disease heterogeneity, the population size for each etiology in our cohort remains modest. It is known that there is a topography difference in fibrosis deposition between NASH and viral hepatitis. Model inferences in several validation cohorts including only NASH and only viral hepatitis are mandatory to evaluate models' stabilities. Recognizing the inherent challenges associated with dataset scale, deliberate adjustments were undertaken for model hyperparameters, encompassing the minimum number of trees, maximum tree layers, and branch reduction within the context of the random forest framework. These methodical interventions were orchestrated to preserve model simplicity while concurrently avoiding overfitting problems. Third, the MRI protocol acquisition used in the study is not conventional, particularly for the use of chemical shift encoded MRI and multiple b-value acquisition. We acknowledge that it may constitute an additional limitation for dissemination purposes and require a standardization initiative. However, to date, CSE-MRI and multiple b-values pulse sequences are available in numerous MR systems.

TABLE I

DIFFUSION-BASED MRI MODEL'S PARAMETERS AND THEIR FUNCTION

| | Parameters | Function |
|---|---|---|
| **MONO** | ADC ($mm^2/s$) | $\mathbf{S}_i = \mathbf{S}_0 exp(-\mathbf{b} \times ADC)$ |
| **Stat_D** | Ds ($mm^2/s$)<br>Σ ($mm^2/s$) | $\mathbf{S}_i = \mathbf{S}_0 exp(-\mathbf{b} \times D_s + \frac{1}{2} \times \mathbf{b}^2 \sigma^2)$ |
| **SEM** | DDC ($mm^2/s$)<br>α | $\mathbf{S}_i = \mathbf{S}_0 exp[-(\mathbf{b} \times D)^{\alpha}]$ |
| **IVIM** | f<br>D ($mm^2/s$)<br>D* ($mm^2/s$) | $\mathbf{S}_i = \mathbf{S}_0 [(1-f) exp(-\mathbf{b} \times D) + f \times exp(-\mathbf{b} \times D^*)]$ |

TABLE II

CLINICAL CHARACTERISTICS OF PATIENTS IN TRAINING AND TESTING SETS BY FIBROSIS AND INFLAMMATION

| Target | Characteristics | Mild | Severe | p-value |
|---|---|---|---|---|
| | **Sex** | | | * |
| | Male | 18 | 26 | |
| | Female | 23 | 10 | |
| | **Age** | | | 0.976 |
| | Mean | 48.85 | 48.89 | |
| | Std | 13.85 | 16.32 | |
| | BMI | | | 0.68 |
| | Mean | 27.44 | 27.85 | |
| | Std | 5.58 | 5.85 | |
| Fibrosis | **Clinical etiology** | | | |
| | NASH | 7 | 12 | 0.12 |
| | Hepatitis B | 11 | 9 | 1.00 |
| | Hepatitis C | 6 | 11 | 0.11 |
| | **Others** | | | |
| | Alcohol | 1 | 2 | 0.60 |
| | Liver transplant | 6 | 1 | 0.11 |
| | **Inflammation** | | | ** |
| | No | 18 | 5 | |
| | A1-A2 | 23 | 31 | |
| | **Sex** | | | 0.80 |
| | Male | 14 | 30 | |
| | Female | 9 | 24 | |
| | **Age** | | | 0.49 |
| | Mean | 50.17 | 48.31 | |
| | Std | 15.79 | 14.70 | |
| | BMI | | | 0.33 |
| | Mean | 26.66 | 28.9 | |
| | Std | 5.53 | 5.74 | |
| Inflammation | **Clinical etiology** | | | |
| | NASH | 6 | 13 | 1.00 |
| | Hepatitis B | 4 | 16 | * |
| | Hepatitis C | 5 | 13 | * |
| | **Others** | | | |
| | Alcohol | 1 | 2 | |
| | Liver transplant | 3 | 4 | 1.00 |
| | **Fibrosis (Ishak)** | | | ** |
| | F0-F2 | 18 | 23 | |
| | F3-F6 | 5 | 31 | |

The student t-test was used to compare the difference in age and BMI and Fisher's exact test was used to compare the difference in other characters. (*$P < 0.05$; **$P < 0.01$)

TABLE III PERFORMANCE OF SINGLE SEQUENCE MODEL

| target | cohort | Sequence | AUC(%) | Acc(%) | Preci(%) | Sen(%) | Spe(%) | F1 | Cut-off(%) |
|---|---|---|---|---|---|---|---|---|---|
| **Inflammation** | Training | T1WI | 95 (93-97) | 90 (87-94) | 93 (88-100) | 88 (82-95) | 93 (87-100) | 90 (88-94) | 52 (44-60) |
| | | T2WI | 97 (96-99) | 94 (90-95) | 93 (87-97) | 94 (90-95) | 93 (87-97) | 93 (86-98) | 51 (40-58) |
| | | CSE(PDFF) | **99 (99-100)** | 97(95-100) | 95 (91-100) | **99 (97-100)** | 95(90-100) | 97 (95-100) | 41 (33-45) |
| | | Diffusion (DDC+σ+D+D*) | 99 (97-100) | 97 (95-100) | **97 (93-100)** | 97 (93-100) | **97 (92-100)** | 97 (95-100) | 52 (45-57) |
| | Test | T1WI | 73 (66-83) | 68 (59-74) | 69 (58-77) | 73 (64-79) | 63 (38-77) | 70 (67-74) | - |
| | | T2WI | 82 (73-86) | 74 (62-79) | 78 (58-100) | 71 (50-79) | **77 (47-100)** | 73 (67-79) | - |
| | | CSE(PDFF) | 85 (78-94) | 76 (64-86) | 76 (62-92) | 79 (71-86) | 73 (57-93) | 77 (67-85) | - |
| | | Diffusion (DDC+σ+D+D*) | **93 (89-96)** | **82(69-88)** | **78 (62-92)** | **94 (85-100)** | 71 (38-92) | **85 (76-89)** | - |
| **Fibrosis** | Training | T1WI | 97 (94-99) | 94 (89-96) | 93 (89-100) | 94 (89-100) | 93 (90-100) | 93 (89-96) | 51 (40-63) |
| | | T2WI | 94 (90-97) | 90 (88-95) | 90 (84-96) | 90 (81-96) | 91 (82-97) | 90 (86-94) | 51 (42-60) |
| | | CSE(PDFF+T2*) | 98 (95-99) | **96 (95-98)** | **96 (93-100)** | 95 (93-96) | **97 (93-100)** | **96 (94-98)** | 51 (42-60) |
| | | Diffusion (DDC+σ+D+D*) | **99 (98-99)** | 95 (93-98) | 94 (87-100) | 97 (93-100) | 94 (87-1) | 95 (93-98) | 48 (39-56) |
| | Test | T1WI | 35 (25-46) | 41 (30-60) | 29 (0-55) | 33 (0-67) | 47 (27-64) | 30 (0-60) | - |

| | | | | | | | |
|---|---|---|---|---|---|---|---|
| T2WI | 61 (49-68) | 59 (50-70) | 55 (40-67) | 56 (22-78) | 62 (27-82) | 53 (29-67) | - |
| CSE(PDFF) | 79 (68-92) | **75 (65-90)** | **79 (58-100)** | 64 (44-89) | **84 (55-100)** | 69 (53-89) | - |
| Diffusion (DDC+σ+D+D*) | **82 (77-90)** | 71 (65-80) | 75 (57-100) | **67 (33-89)** | 75 (45-100) | 66 (50-80) | - |

TABLE IV THE PERFORMANCE OF MULTIPLE SEQUENCES

| target | cohort | Sequence | AUC(%) | Acc(%) | Preci(%) | Sen(%) | Spe(%) | F1 | Cut-off(%) |
|---|---|---|---|---|---|---|---|---|---|
| **Inflammation** | Training | Two sequences (CSE+Diffusion) | 100 (100-100) | 98 (96-100) | 99 (97-100) | 98 (93-100) | 98 (97-100) | 98 (96-100) | 54 (42-65) |
| | | Three sequences (T2+CSE+Diffusion) | 99 (98-100) | 96 (93-98) | 92 (87-97) | 98 (95-100) | 95 (86-98) | 96 (93-98) | 50 (35-70) |
| | | Four sequences (T2+CSE+ Diffusion+Inflam) | 100 (100-100) | **99 (99-100)** | **100 (98-100)** | **99 (97-100)** | **100 (98-100)** | **99 (99-100)** | 54 (45-65) |
| | Test | Two sequences (CSE+Diffusion) | 98 (96-100) | 91 (81-96) | 94 (87-100) | 89 (71-100) | 94(85-100) | 91 (80-97) | - |
| | | Three sequences (T2+CSE+Diffusion) | 96 (94-97) | 84 (79-96) | 85 (70-100) | 87 (57-100) | 80 (57-100) | 84 (73-96) | - |
| | | Four sequences (T2+CSE+ Diffusion+Inflam) | **100 (98-100)** | **96 (89-100)** | **95 (82-100)** | **99 (93-100)** | 94 (79-100) | **97(90-100)** | - |
| **Fibrosis** | Training | Two sequences (CSE+Diffusion) | 99 (98-100) | **97 (95-98)** | 94 (90-96) | 99 (96-100) | 95 (90-97) | **97 (95-98)** | 45 (39-51) |
| | | Three sequences (T2+CSE+Diffusion) | 99 (99-100) | 95 (93-96) | 92 (87-93) | 99 (96-100) | 92 (87-93) | 95 (93-96) | 42 (37-50) |
| | | Four sequences (T2+CSE +Diffusion+Inflam) | 99 (98-100) | 96 (93-100) | **95 (87-100)** | 98 (89-100) | 95 (87-100) | 96 (93-100) | 48 (40-55) |
| | Test | Two sequences (CSE+Diffusion) | 85 (78-90) | 76 (65-85) | 69 (56-75) | 89 (67-100) | 65 (36-82) | 77 (71-86) | - |
| | | Three sequences (T2+CSE+Diffusion) | 89 (80-95) | 81 (75-90) | 73 (67-82) | **91 (78-100)** | 73 (64-82) | 81 (74-90) | - |
| | | Four sequences (T2+CSE+ Diffusion+Inflam) | **97 (94-100)** | **83 (75-90)** | **85 (67-100)** | 80 (67-89) | **85 (64-100)** | 81 (76-88) | - |

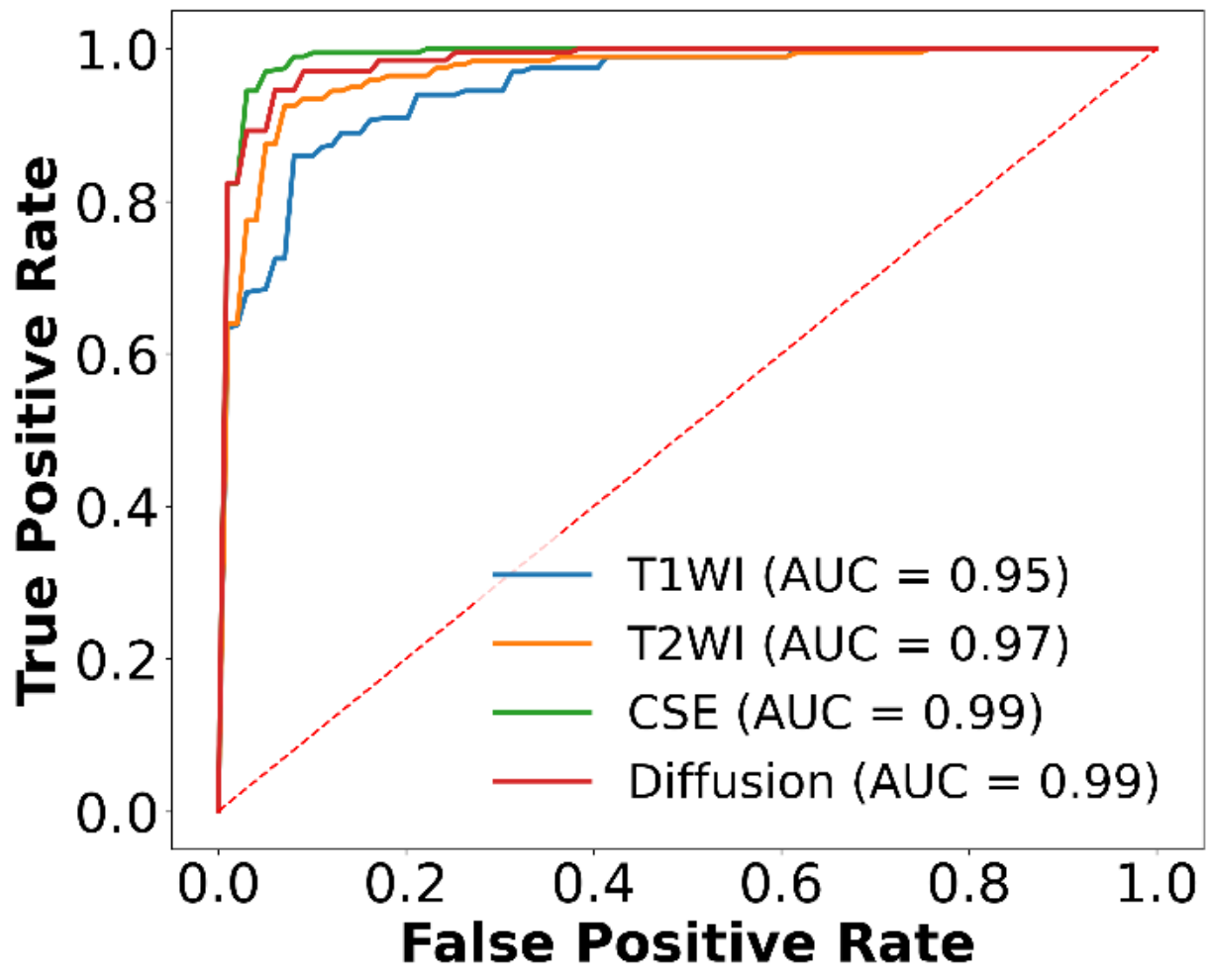

(a)

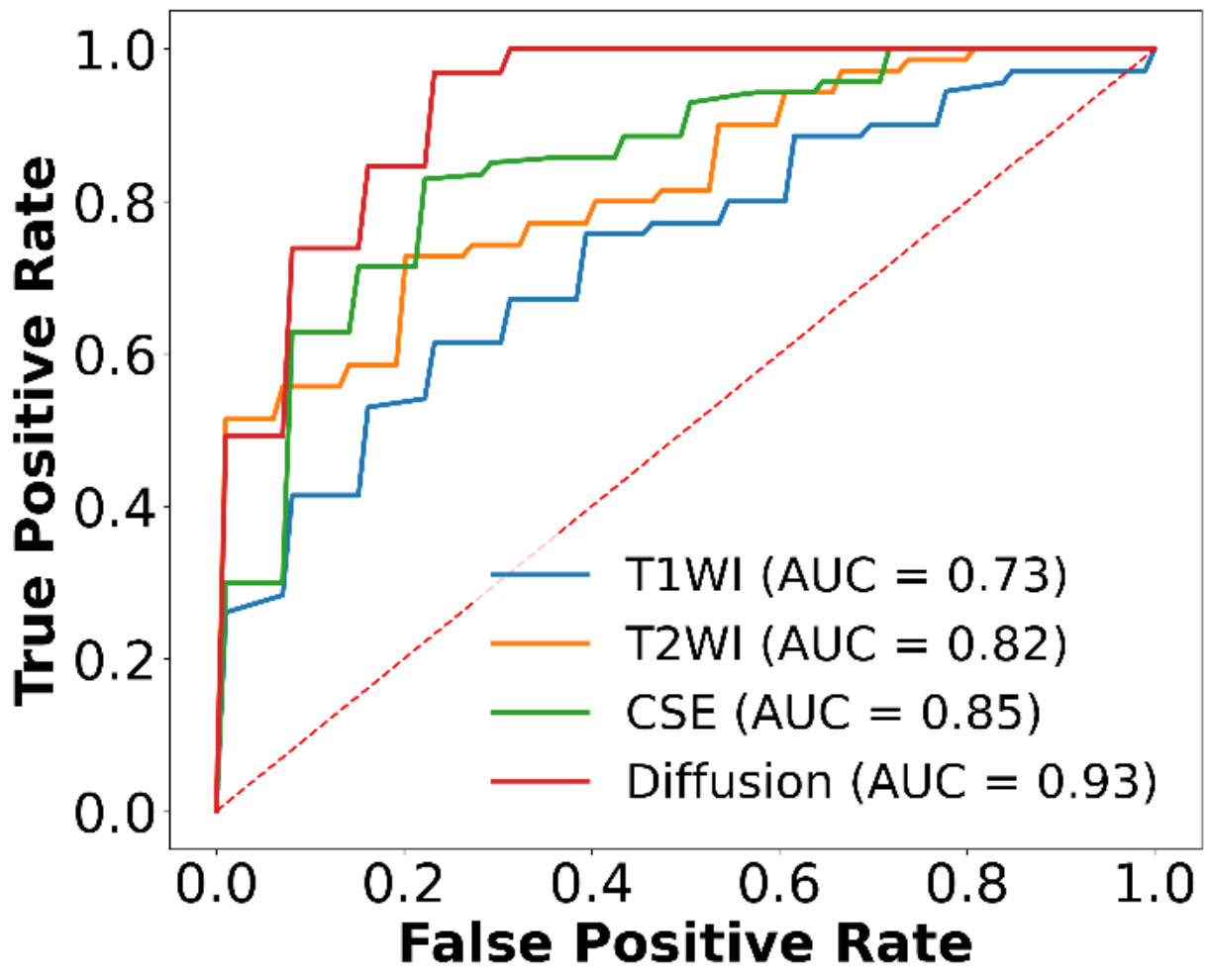

(b)

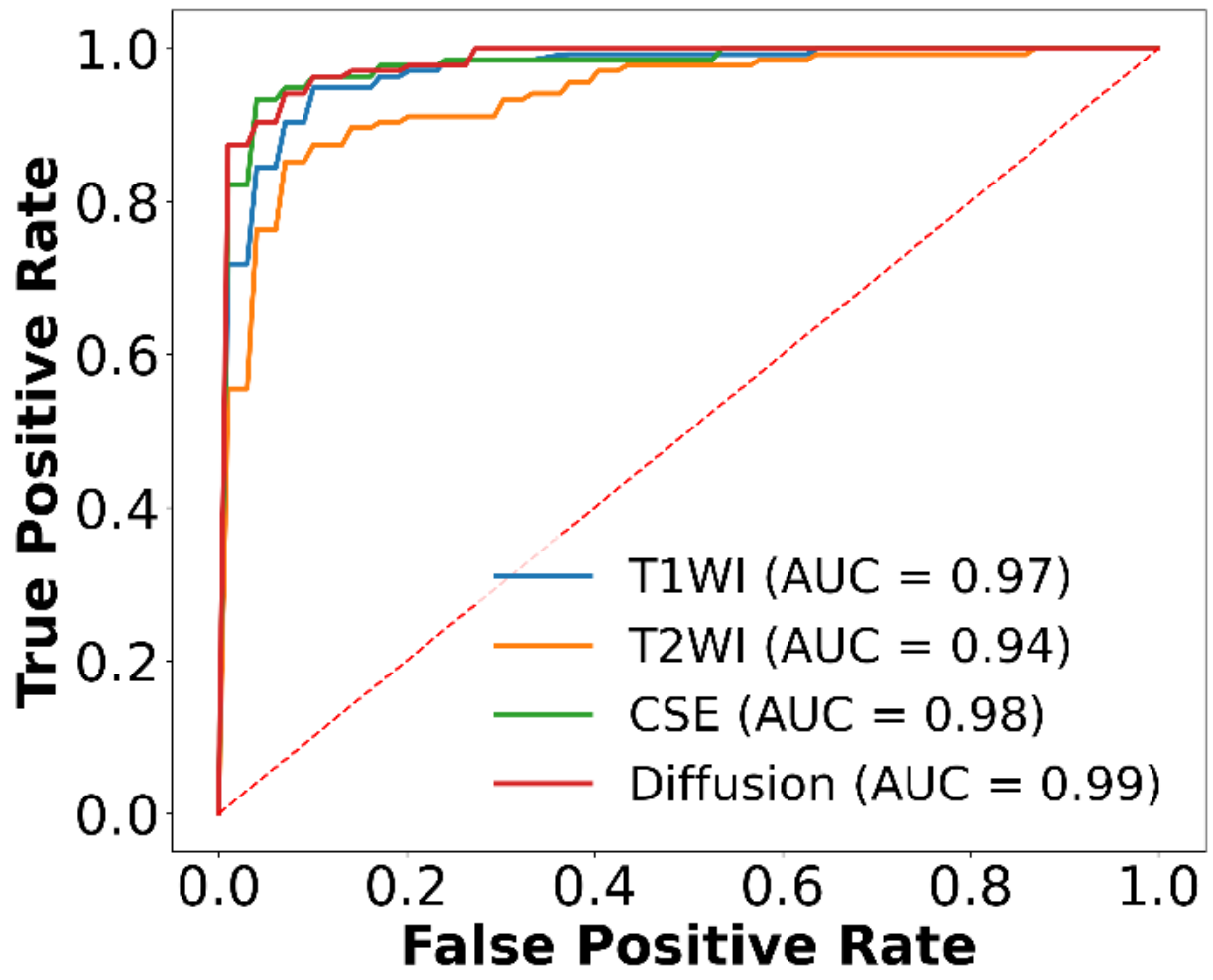

(c)

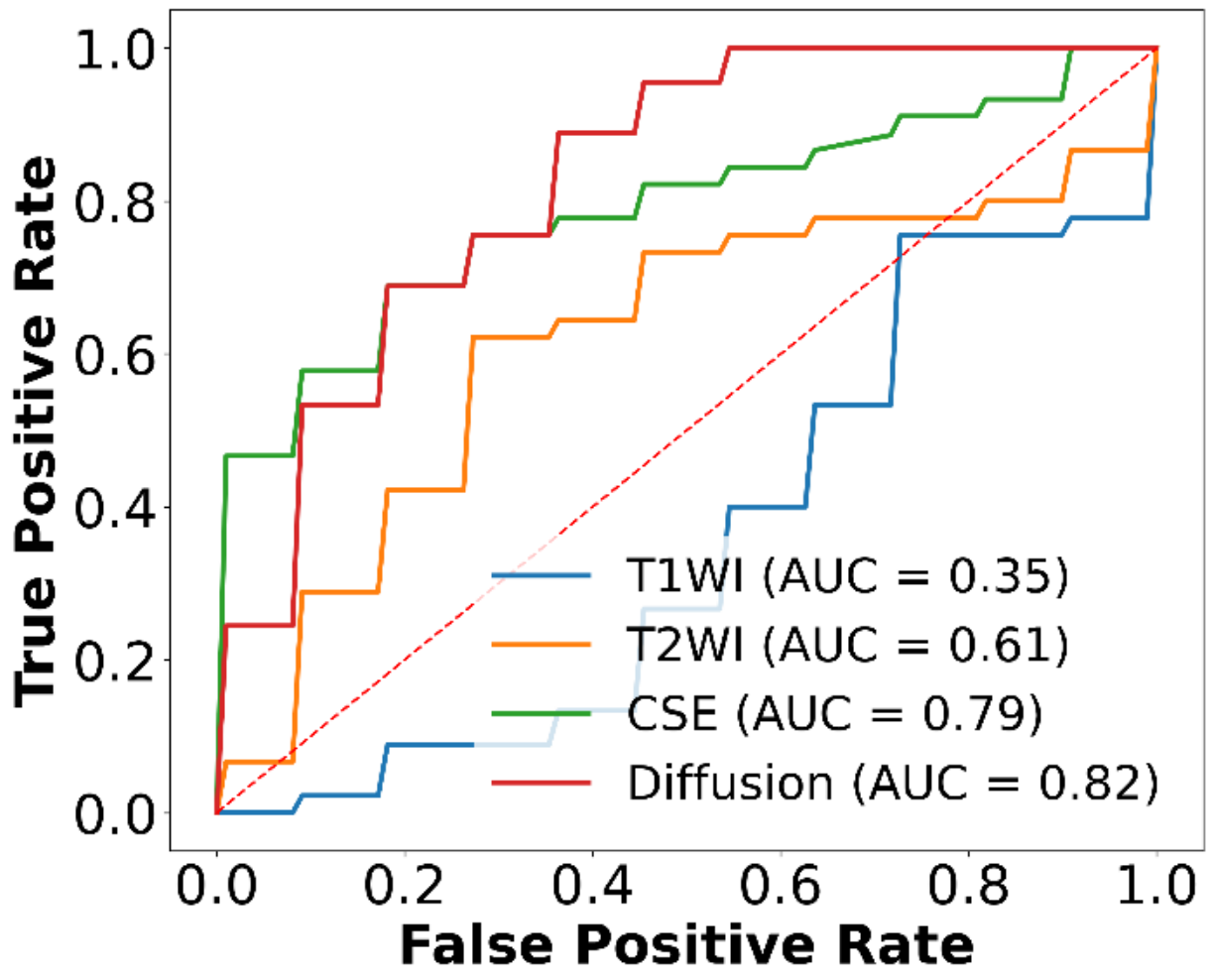

(d)

**Fig. 1 Comparison of ROC curves using a single sequence for the classification. (a) and (b) are ROC curves in case of inflammation classification for training and test set. (c) and (d) are ROC curves in case of fibrosis classification for the training and test set.**

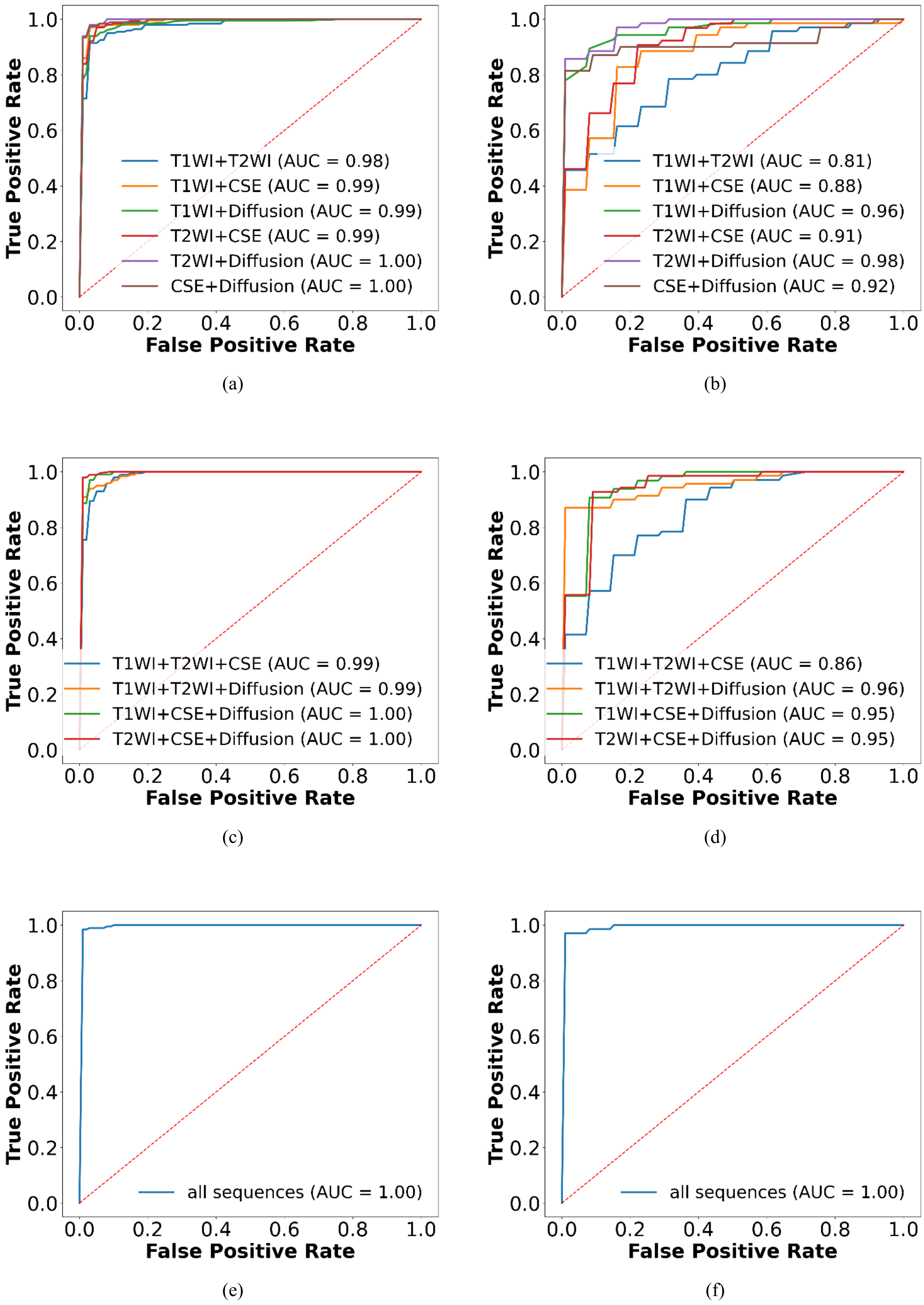

**Fig.2 Comparison of ROC curves of multiple sequences for inflammation classification.**

**(a) and (b) are ROC curves with two sequences in the training and test set. (c) and (d) are ROC curves with triple sequences in the training and test set. (e) and (f) are ROC curves with all sequences in the training and test set.**

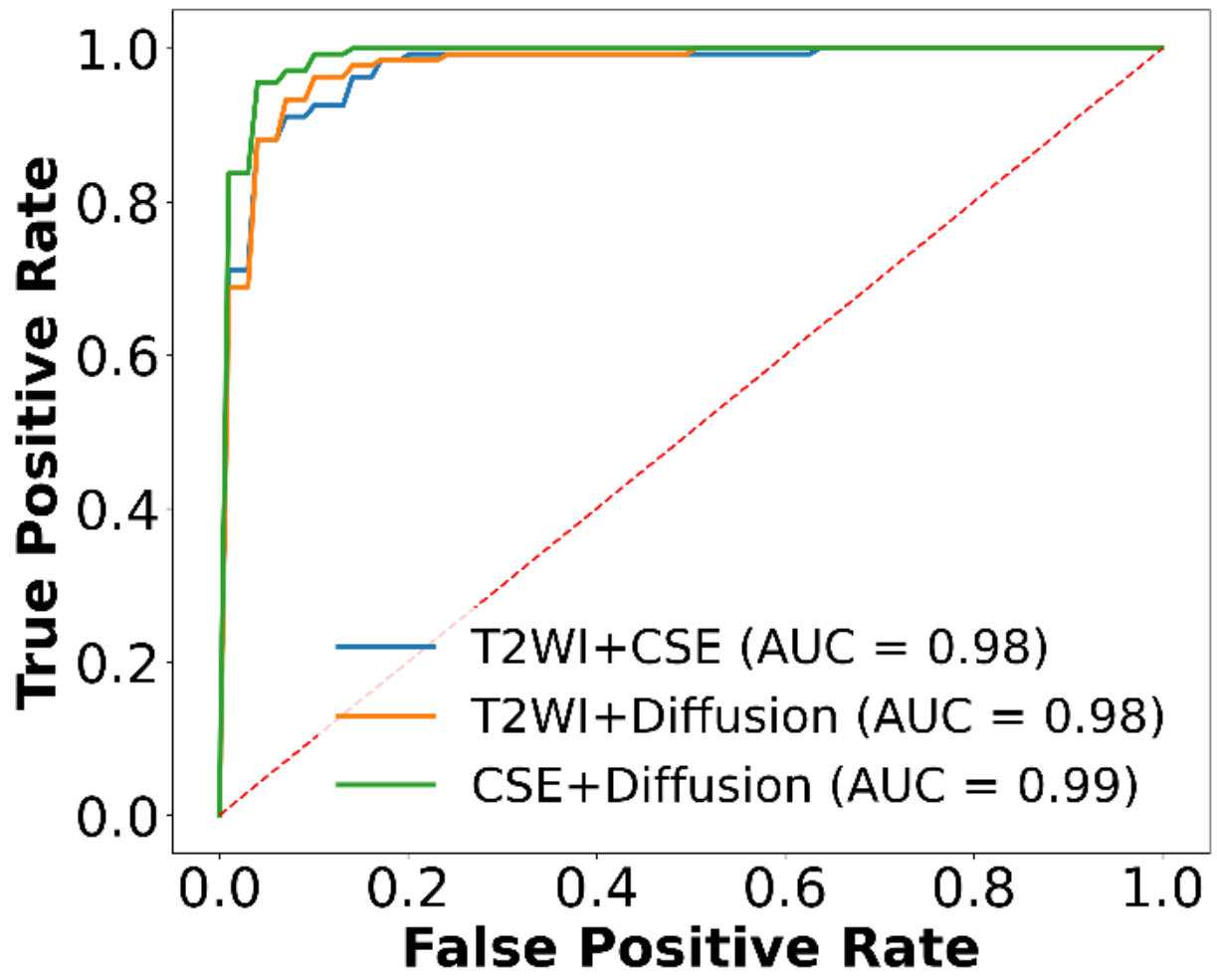

(a)

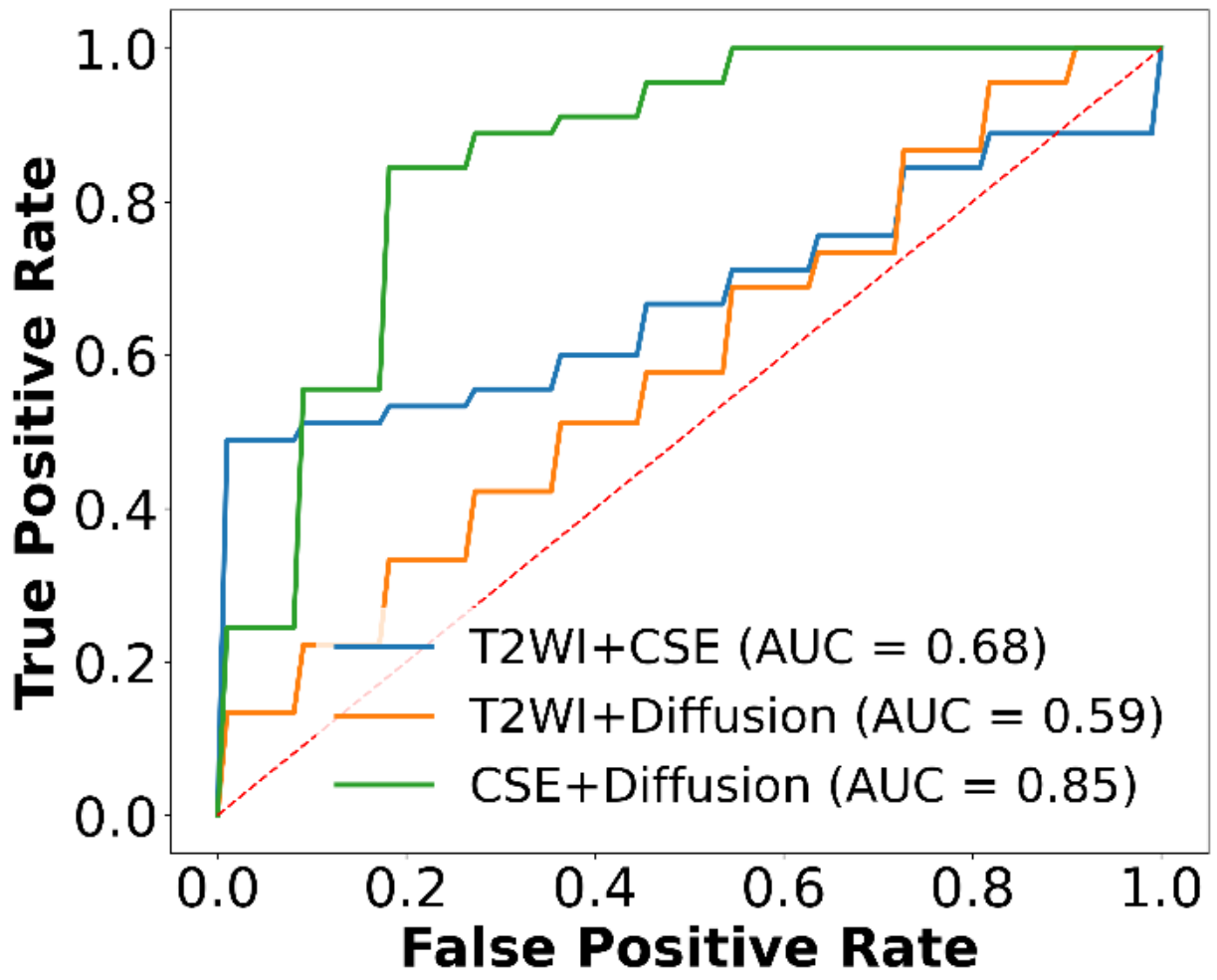

(b)

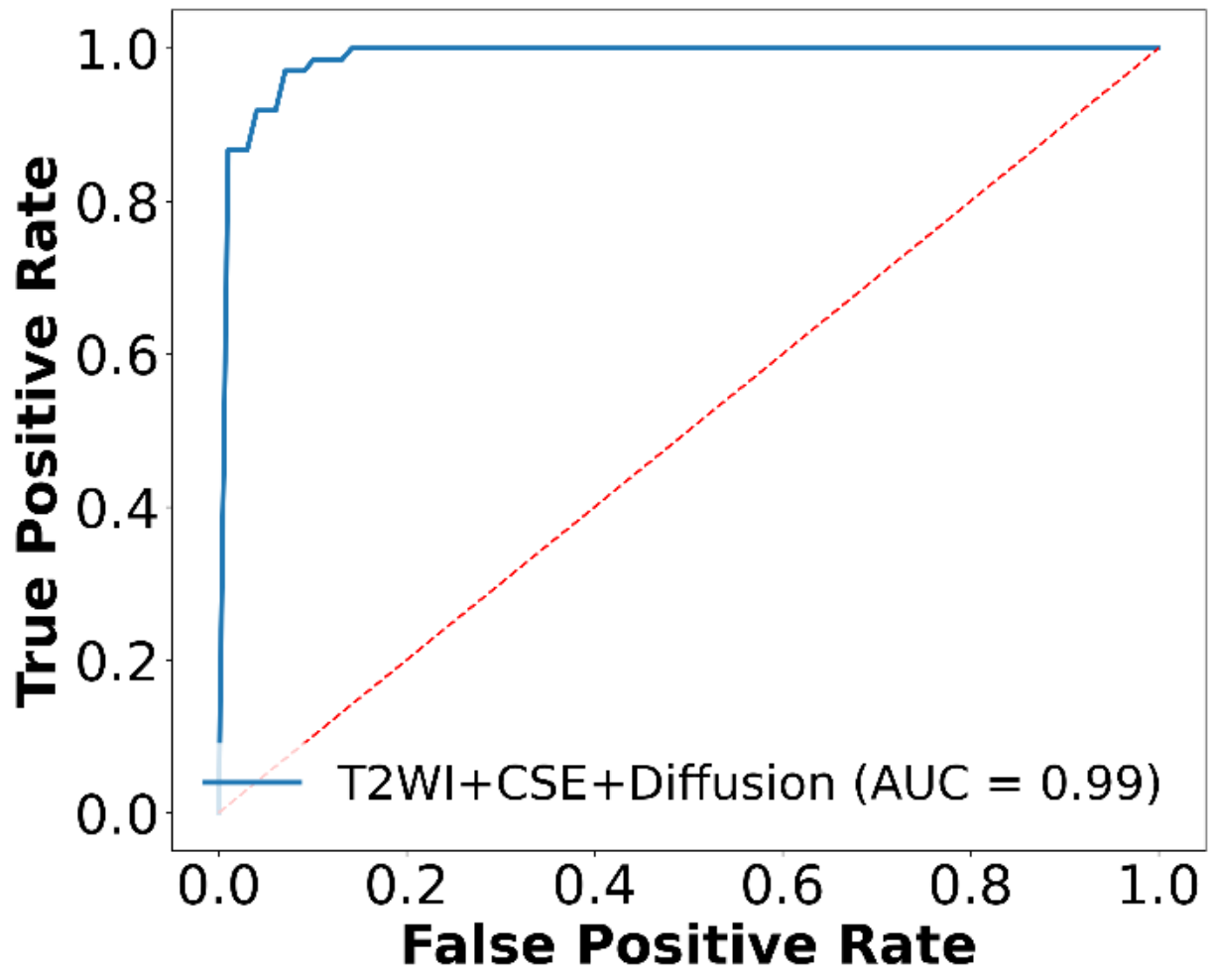

(c)

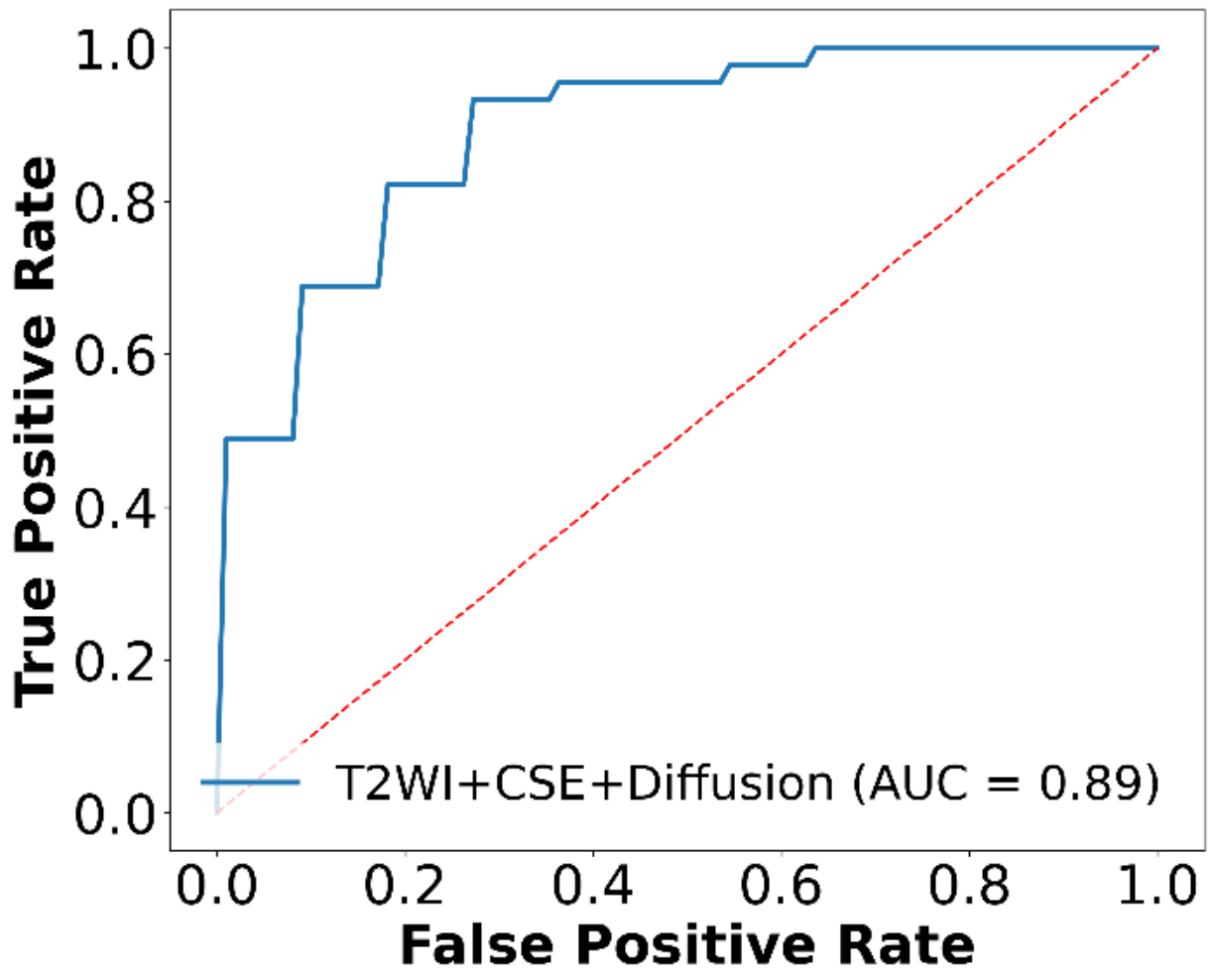

(d)

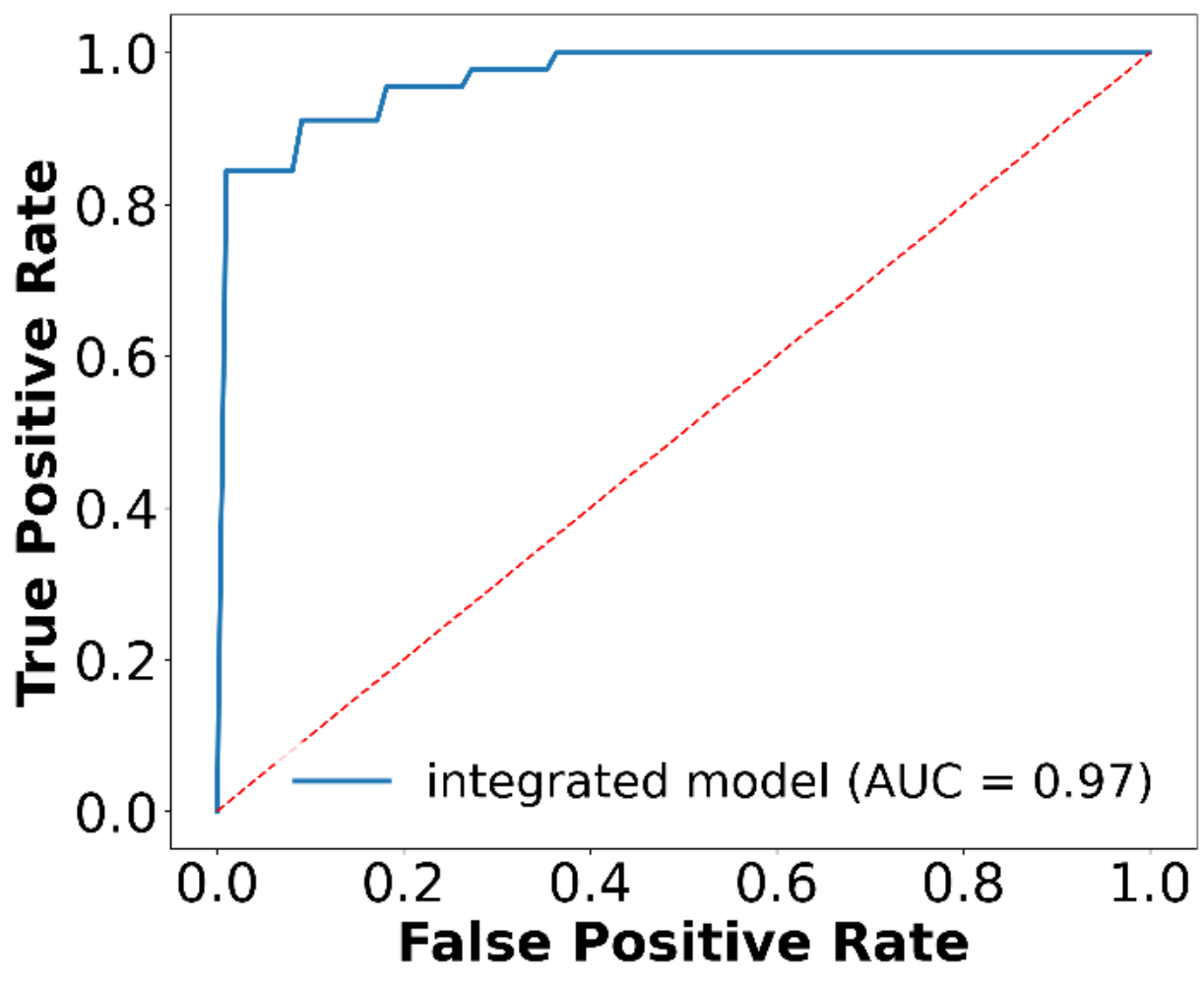

(e)

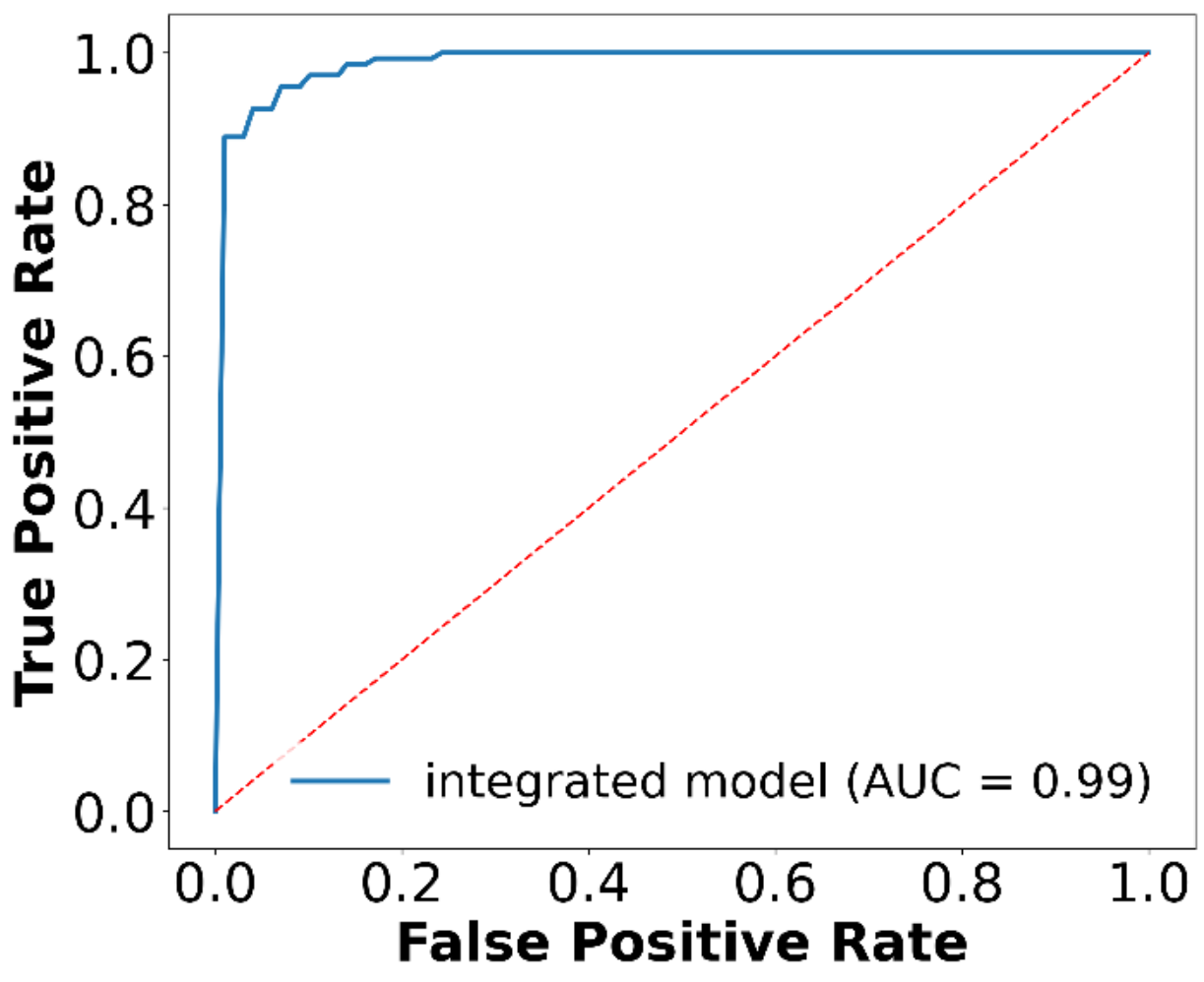

(f)

**Fig.3 Comparison of ROC curves of multiple sequences for fibrosis classification. (a) and (b) are ROC curves with two sequences in the training and test set. (c) and (d) are ROC curves with triple sequences in the training and test set. (e) and (f) are ROC curves with triple sequences and inflammation prediction scores in training and test set.**

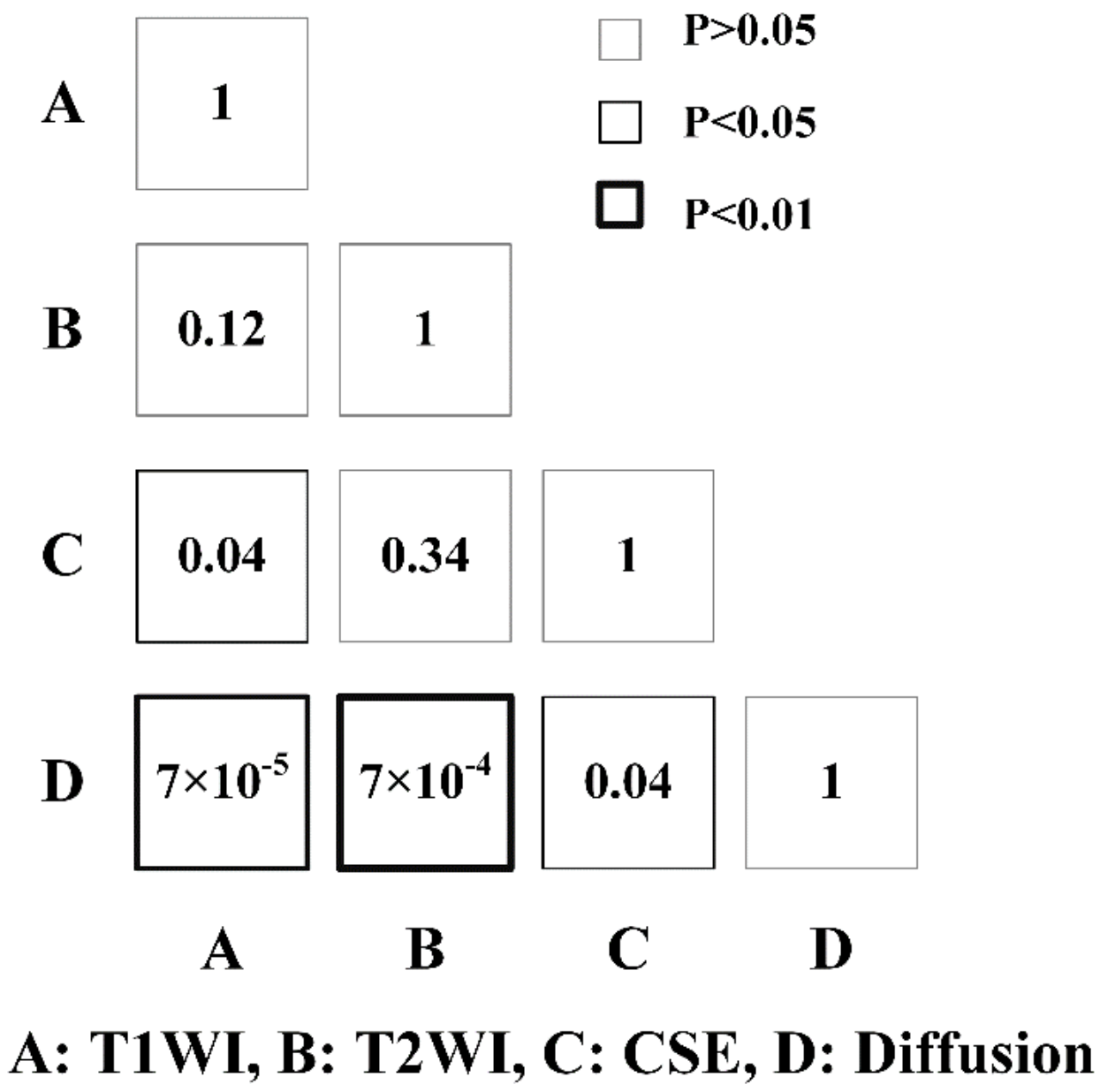

(a)

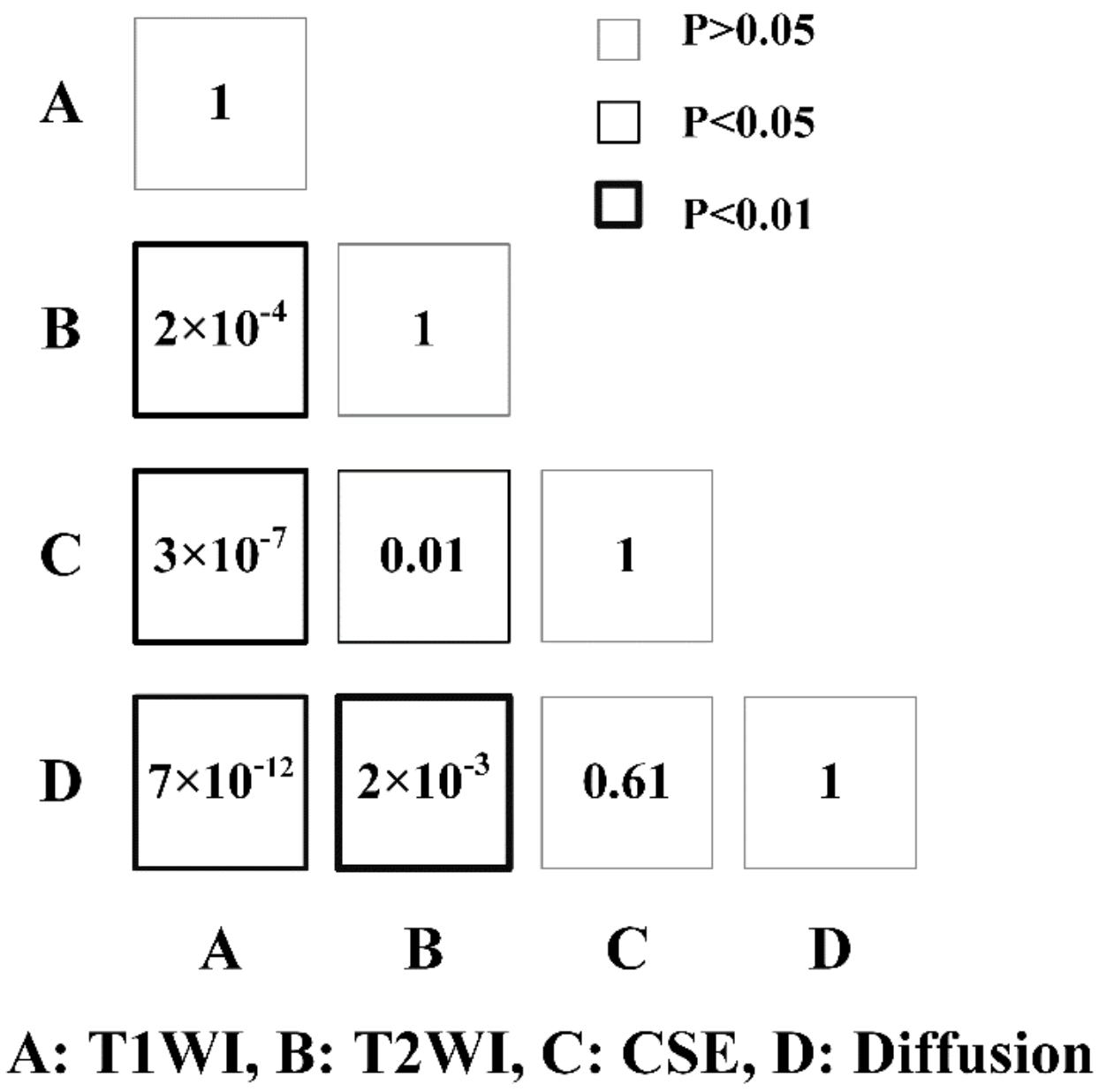

(b)

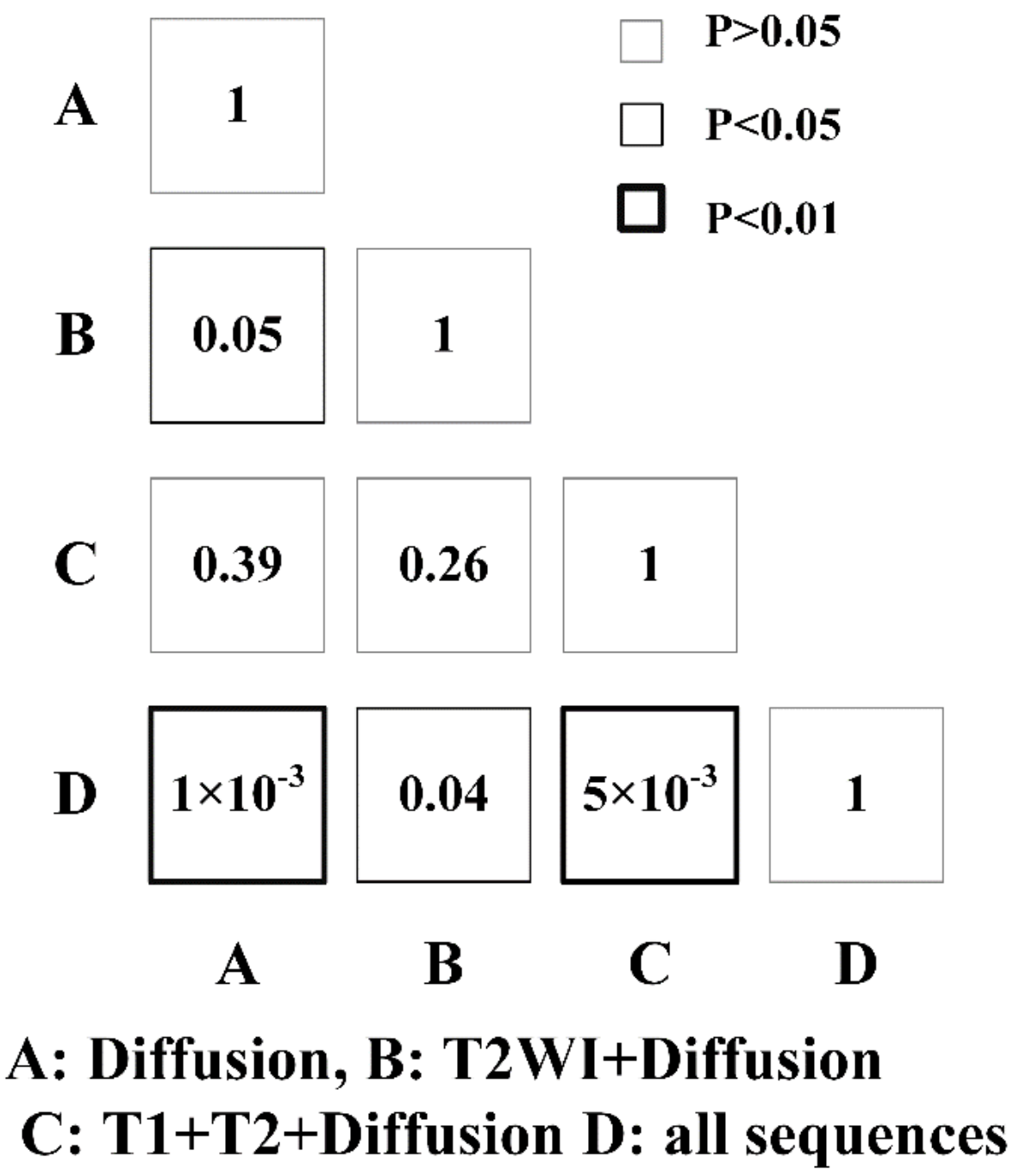

(c)

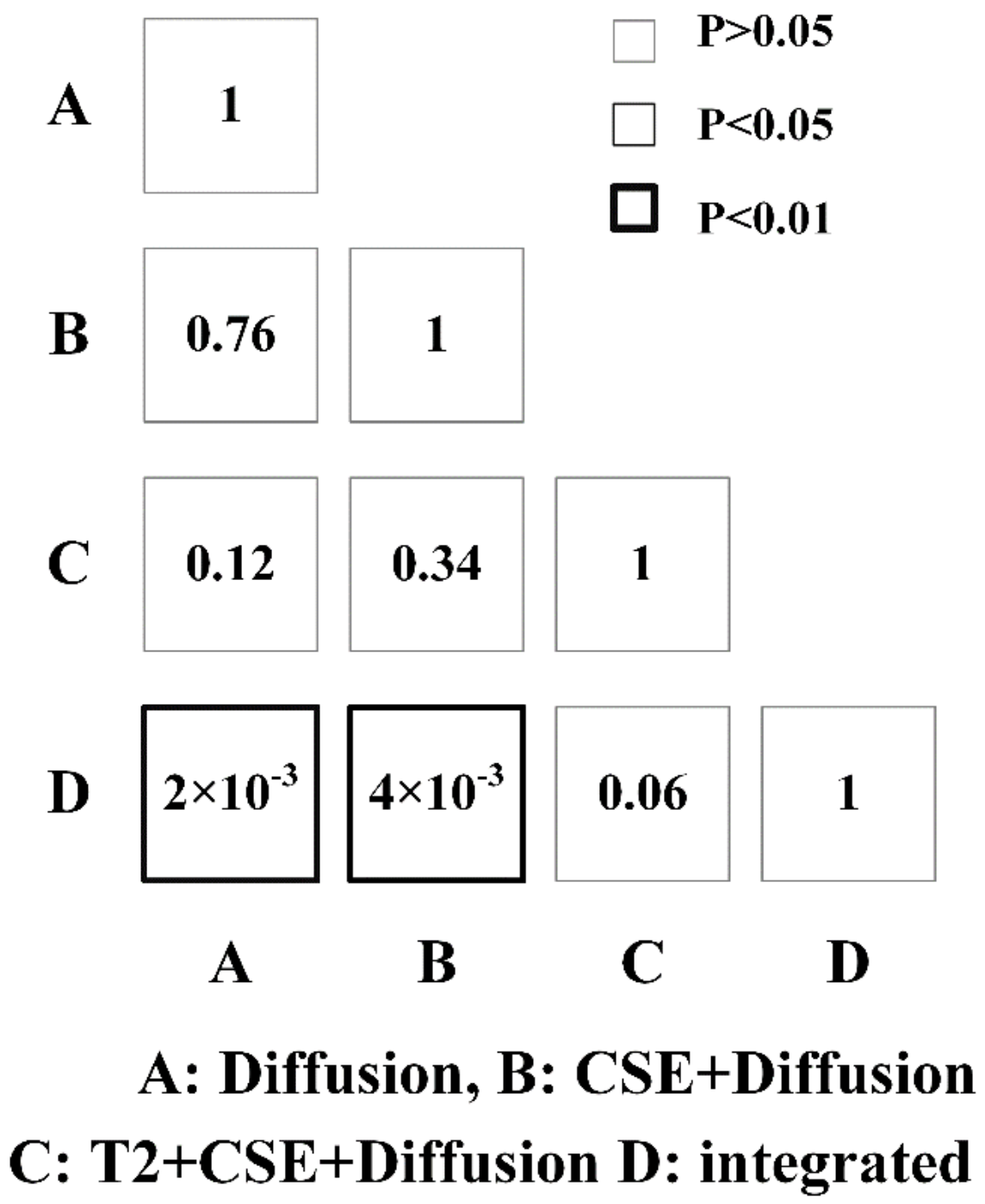

(d)

**Fig.4. Delong test between different models. (a) and (b) are the Delong test p-values between ROC for inflammation and fibrosis classification models trained from radiomics extracted from a single MRI sequence. (c) and (d) are the Delong test p-values between ROC for inflammation and fibrosis classification models trained with radiomics extracted from multiple sequences.**